\documentclass[%
superscriptaddress,
preprint,
 amsmath,amssymb,
 aps,
]{revtex4-1}

\usepackage{graphicx}% Include figure files
\usepackage{dcolumn}% Align table columns on decimal point
\usepackage{subfigure}
\usepackage{bm}% bold math
\usepackage{threeparttable}
\usepackage{multirow}

\begin{document}

\title{An exponent tunable network model for reproducing density driven superlinear relation}% Force line breaks with \\

\author{Yuhao Qin}
%  \altaffiliation[Also at ]{Physics Department, XYZ Universality.}%Lines break automatically or can be forced with \\
\affiliation{%
 Systems Science Institute, State Key Laboratory of Rail Traffic Control and Safety, MOE Key Laboratory for Urban Transportation Complex Systems Theory and Technology, Beijing Jiaotong University, Beijing 100044, China
}%
\author{Liang Gao}%
\email{Corresponding Author. email: lianggao@bjtu.edu.cn}
\affiliation{%
 Systems Science Institute, State Key Laboratory of Rail Traffic Control and Safety, MOE Key Laboratory for Urban Transportation Complex Systems Theory and Technology, Beijing Jiaotong University, Beijing 100044, China
}%
\author{Lida Xu}
\email{Corresponding Author. email: xuld@buct.edu.cn}
\affiliation{%
 Beijing Key Laboratory of Bioprocess, College of Life Science and Technology, Beijing University of Chemical Technology, Beijing 100029, China
}%
\author{Zi-You Gao}
\affiliation{%
 Systems Science Institute, State Key Laboratory of Rail Traffic Control and Safety, MOE Key Laboratory for Urban Transportation Complex Systems Theory and Technology, Beijing Jiaotong University, Beijing 100044, China
}%
\date{\today}% It is always \today, today,

\begin{abstract}
Previous works have shown the universality of allometric scalings under density and total value at city level, but our understanding about the size effects of regions on them is still poor. Here, we revisit the scaling relations between gross domestic production (GDP) and population (POP) under total and density value. We first reveal that the superlinear scaling is a general feature under density value crossing different regions. The scaling exponent $\beta$ under density value falls into the range $(1.0, 2.0]$, which unexpectedly goes beyond the range observed by Pan et al. (Nat. Commun. vol. 4, p. 1961 (2013)). To deal with the wider range, we propose a network model based on 2D lattice space with the spatial correlation factor $\alpha$ as parameter.
Numerical experiments prove that the generated scaling exponent $\beta$ in our model is fully tunable by the spatial correlation factor $\alpha$. We conjecture that our model provides a general platform for extensive urban and regional studies.
\end{abstract}
\maketitle

%\tableofcontents

\section{Introduction}\label{sec:Intro}
Uncovering general patterns of complex systems is a major task for science. Since individuals behave differently in complex systems, statistical properties are usually chosen to investigate the potential patterns of a complex system. The scaling law is one of the statistical properties recently observed in regional systems \cite{Bettencourt2010,Bettencourt2007,Shalizi2011,Ortman2014,Arcaute2014}.

Our world now contains more people than any time in human history \cite{Crane2005}. Human's activities exert a significant influence on the growth of socioeconomic indicators of a regional system \cite{Gonzalez2008}. Most of these indicators ($Y$), related to the productivity and creation \cite{Florida2002}, are determined by the superlinear scaling law: $Y=c*N^\beta$, where $N$ is the population size of a region, $c$ is a normalization constant and $\beta$ is the scaling exponent larger than 1 \cite{Bettencourt2007a,Bettencourt2010a,Bettencourt2013,Arcaute2014}.
(see SI, Sec. S2).

To understand the mechanism of the superlinear scaling in regional systems, several network models have been established. Arbesman et al. built a first network model with hierarchical tree structure to reproduce the superlinear scaling \cite{Arbesman2009}, but their assumption is hard to be found in real systems \cite{Ahn2010,Mucha2010,Onnela2011,Expert2011}. Bettencourt developed another network model according to human mobility which can fit the empirical data well \cite{Bettencourt2013}, but how the social structure of his model influences the superlinear scaling has not been explained clearly. To improve Bettencourt model, Yahubo et al. introduced geographic features into network model, and provided a possible explanation for the influence of social structure on the superlinear scaling \cite{Yakubo2014}.

More recently, Pan et al. discovered superlinear relations in country level between per area socioeconomic indicators (GDP, number of patents, etc.) and population density \cite{Pan2013}. These superlinear relations under density value are stronger than those under total value. To explain the superlinear scaling under density value, they established a rank-based network model by attributing the interplay among geography, population and societal interaction to density-driven tie formation. Despite the simulation results well agreeing with the empirical data, the superlinear scaling exponent $\beta$, generated by their model, is limited in a narrow range [1.1, 1.3]. Actually, a superlinear relation exists in different regional levels (world, continent, country, region of a country, province and city) and its exponents $\beta$ varies within a wider range (1,2] for different region levels \cite{Pumain2006,Bettencourt2007,Arbesman2009,Nomaler2014,Arcaute2014}.

Therefore, we propose here an exponent tunable network model to reproduce the superlinear scaling relation under density value, whose range of the scaling exponent is extended to cover the wider range observed in real world with different regional levels.

Considering the superlinear relation at different regional systems as a consequence of different degree of spatial correlation, which is measured by the parameter $\alpha$, we therefore believe our model offers a generative indication for the development of regional economics without the information of social structure.

\section{Empirical Studies}\label{sec:Emprical}
Scaling relations under total and density values have been observed between socioeconomic indicators and population \cite{Bettencourt2007,Zhang2010,Pan2013,Alves2014}.

\begin{figure} %% Fig.1
\includegraphics[width=\textwidth]{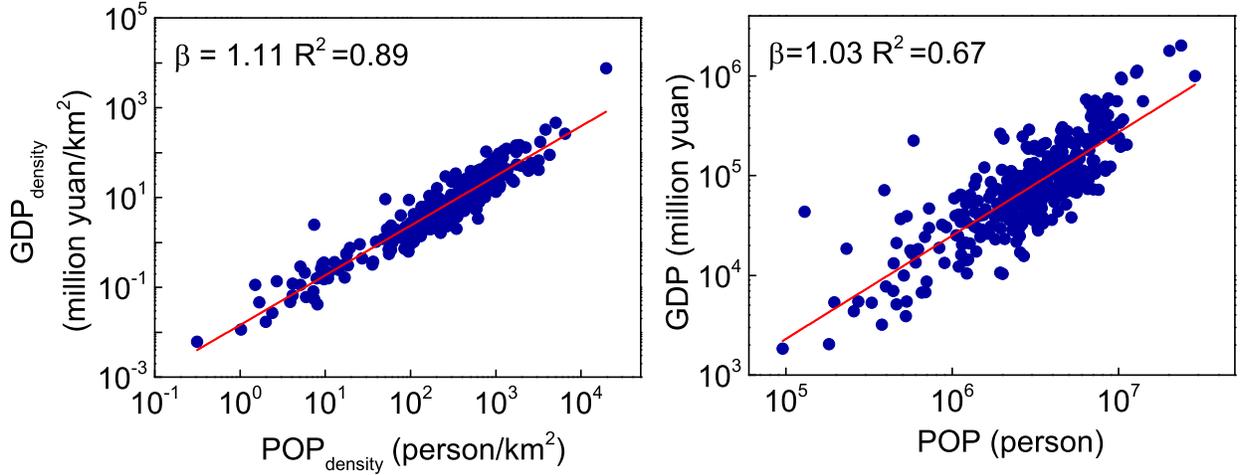}% Here is how to import EPS art
\caption{(a): The rescaled GDP as a function of population density for cities in China. (b): the corresponding plot for GDP as a function of population. In this end of the population scale, we find that density continues to correlate strongly with GDP, whereas the correlation is far less apparent for raw population. This result agrees well with Pan's result.}
\label{fig:China}
\end{figure}

To verify our empirical studies with previous works, we revisit the scaling relations between gross domestic product (GDP) and population (POP) under total and density values. Data for an extensive group of geographic regions are collected from \emph{wikipedia.org} and \emph{Chinese Statistical Yearbook} (see SI, Sec. S1). Fig. \ref{fig:China} and Table \ref{tab:Chinese_Regions} are showing the scaling relations for $GDP_{density}$ vs. $POP_{density}$ and $GDP$ vs. $POP$. According to the adjusted R-square, we find the scaling relations under density value are stronger than that under total value. The scaling exponent $\beta$ under density value for different regions (see Table \ref{tab:Chinese_Regions} and Table \ref{tab:region_levels}) is larger than 1, which is consistent with Pan's results \cite{Pan2013}.

\begin{figure} %% Fig.2
\centering
\includegraphics[width=0.7\textwidth]{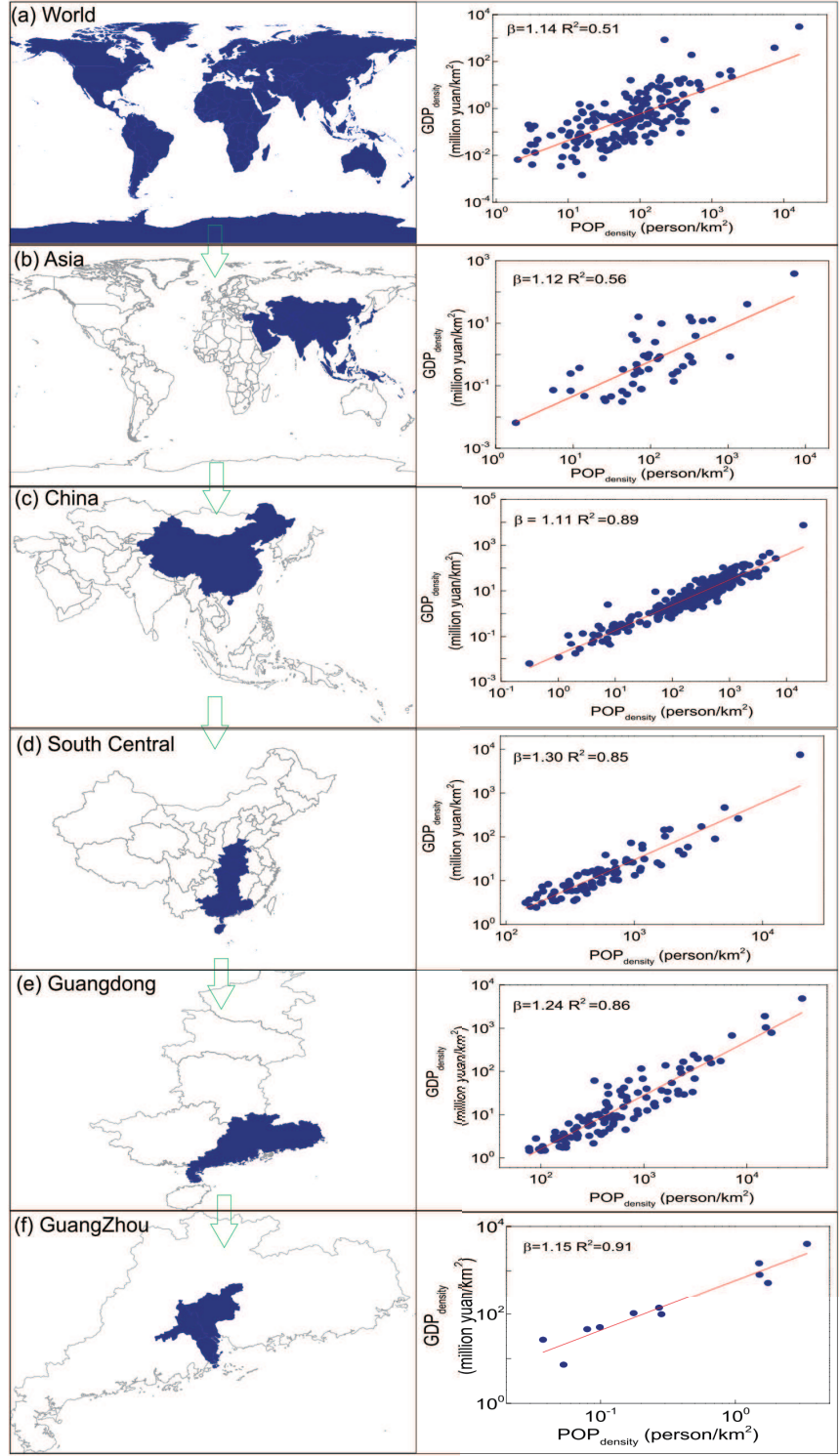}% Here is how to import EPS art
\caption{Scaling relations under denstity value for 6 regional levels. (a) World countries; (b) Asian countries; (c) Chinese cities; (d) Cities in South Central China; (e) Cities in Guangdong province; (f) Counties in Guangzhou metropolis. In all of these pictures, we can see the superlinear relation ($Y\sim X^{\beta}$) under density value at different regional scales. Notice city is a larger administrative area than county in China.}
\label{fig:diffSize}
\end{figure}

To test whether the superlinear scaling under density value is general in different regions, the empirical data of six regional levels are collected (see Table \ref{tab:region_levels}). As shown in Fig. \ref{fig:diffSize}, the superlinear relation between the $GDP_{density}$ and $POP_{density}$ still exists in different regions. Therefore, we conjecture that the superlinear relation under density value is a general feature for different regions.

Despite the general existence of superlinear scaling between the $GDP_{density}$ and $POP_{density}$ in different regions, the superlinear scaling exponents, which are out of the range [1.1, 1.3] reported by Pan \cite{Pan2013}, are also discovered in our empirical studies. For example, all the scaling exponents $\beta$ in bold face in Table \ref{tab:region_levels} and Table \ref{tab:Chinese_Regions} are obviously out of the range in Pan's work. Therefore, an alternative model is required to reproduce the wider range of scaling exponent $\beta$ under density value.

\section{Model Studies}\label{sec:Model}
We propose a growing network model based on a 2D lattice space to reproduce the superlinear relation under density value. In our model, a node represents a small community of individuals, and an edge represents some kind of socioeconomic interactions between individuals in the two nodes. Therefore, the space occupancy $\rho$, which is defined as the number of nodes per unit area, can approximately represent the population density for a real region. And the number of edges ($E$) can approximately represent the socioeconomic product.

Our model starts from an empty lattice of size $L \times L$, with $L^2$ possible locations. At each time step, a new node $j$ is assigned to an empty location on the lattice, and linked to the node(s) already present in the network with the connecting probability $P_{j \rightarrow i}$. Inspired by Kleinberg's network model \cite{Kleinberg2000}, we assume that the connecting probability $P_{j \rightarrow i}$ that a new node $j$ will be connected to a node $i$ is proportion to $r_{ij}$, so that
\begin{equation} \label{eq:con_p}
P_{j \rightarrow i} = \frac{r_{ij}^{-\alpha}}{r_{min}^{-\alpha}}.
\end{equation}
where $r_{ij}$ is the Euclidean distance between nodes $i$ and $j$, $r_{min}$ is the minimum distance among all possible $r_{ij}$ at the present time step, and $\alpha$ is the spatial correlation factor (strongest correlation: $\alpha \to 0$). Eq. (\ref{eq:con_p}) means the farther two nodes apart, the less likely are these two nodes to be connected. Apparently, the space occupancy $\rho$ gradually increases with new nodes added into the lattice space. After a long enough time steps (e.g. $t=10000$ in this paper), the model leads to an exponent tunable and size independent scaling relation between the number of edges $E$ and the space occupancy $\rho$.

\begin{figure} %% Fig.3
\includegraphics[width=\textwidth]{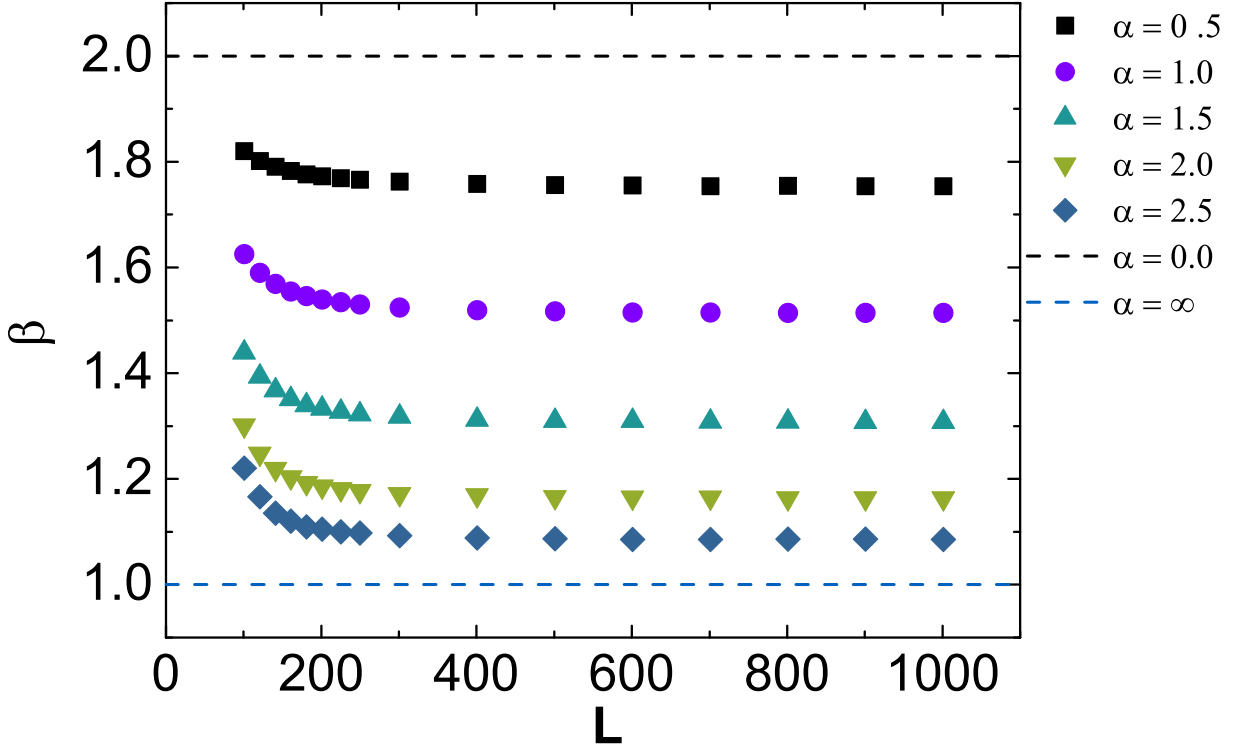}
\caption{The plot of superliner scaling exponent $\beta$ as a function of lattice side length $L$ at different $\alpha$. $\beta$ is averaged over $100$ realizations. Notice with the increase of $L$, $\beta$ quickly converges (e.g. $L = 400$), which means $\beta$ will not be affected scales when $L$ over than $400$. So, to avoid the size effect and system errors in the simulation processes, from here on all simulations are carried on 2D lattice space with $L = 1000$ (The total number of node is $10000$).}
\label{fig:simSizeL}
\end{figure}

Fig. \ref{fig:simSizeL} shows the average over $100$ realizations of the simulation for different values of the linear size $L$ ($100 \leq L \leq 1000$). For each $\alpha$, the scaling exponent $\beta$ quickly converges to a steady value at about $L = 400$. The results proves that when $L \ge 400$, there is no size effect in our model. To avoid the size effect and system errors in the simulation processes, from here on all simulations are carried on 2D lattice space with $L = 1000$ and averaged over $100$ realizations.

\begin{figure} %% Fig.4
\includegraphics[width=\textwidth]{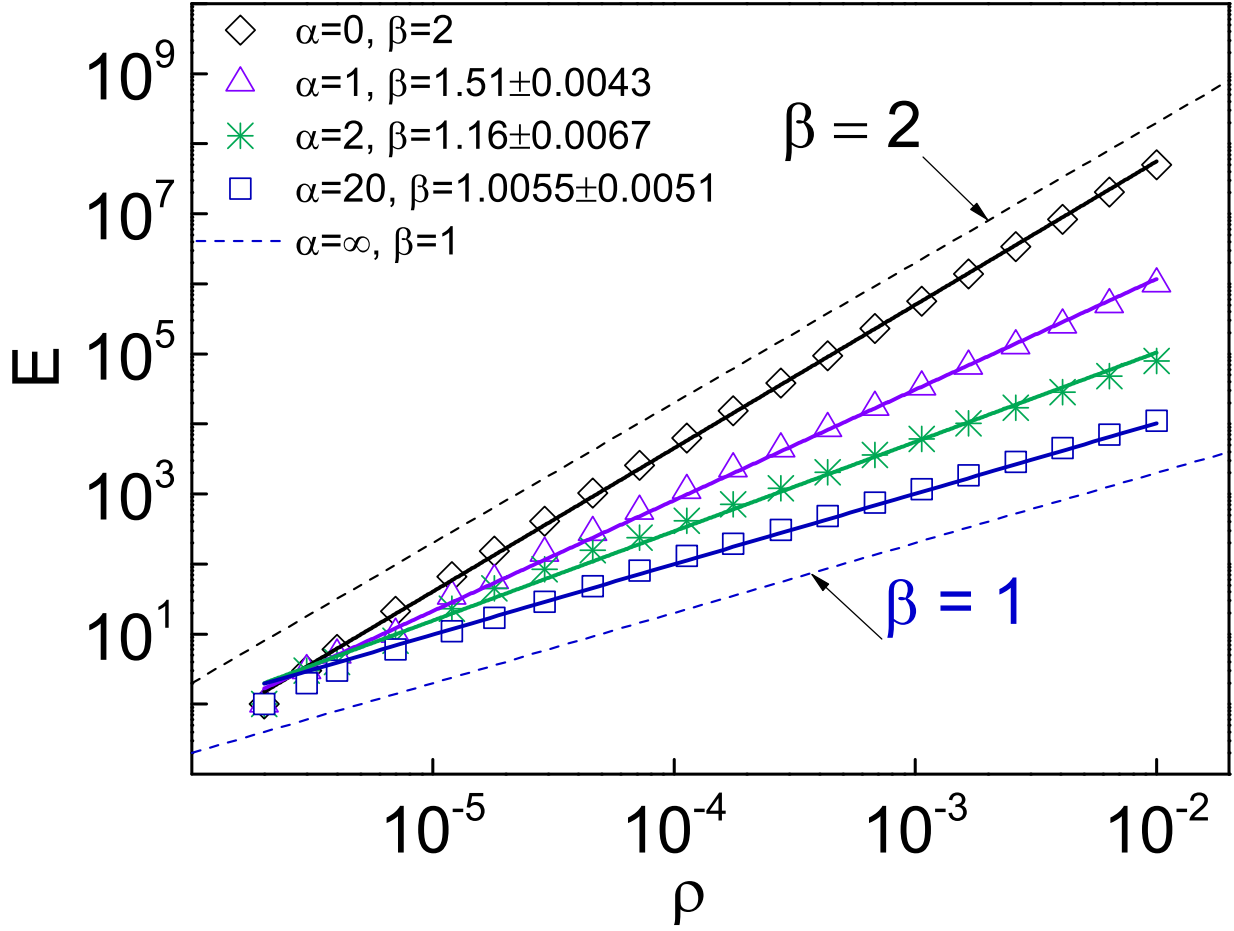}% Here is how to import EPS art
\caption{The plot of the number of edges ($E$) as the function of space occupancy ($\rho$) at different $\alpha$ (the total number of nodes is $10000$). The solid lines are fits to the form $\rho^{\beta}$. For better statistics, we simulate our model at $1000 \times 1000$ for 100 times and take the average value. For each $\alpha$, the agreement is excellent and with the increase of $\alpha$, $\beta$ decreases from $2$ to $1$. This result means our model is reliable to reproduce all superlinear relation under density value. Notice we put the result in logarithmically spaced bins for better transparency.}
\label{fig:space_edge}
\end{figure}

As expected, the network evolves into a stationary state (Fig. \ref{fig:space_edge}), where the number of edges ($E$) grows superlinearly with the space occupancy $\rho$ increasing for different $\alpha$. The scaling exponent $\beta$ decreases from $2.0$ to $1.0$, when $\alpha$ increases from $0.0$ to $\infty$. It indicates that our model can cover a wider scaling range [1.0, 2.0] than Pan's model~\cite{Pan2013}.

\begin{figure} %% Fig.5
\includegraphics[width=\textwidth]{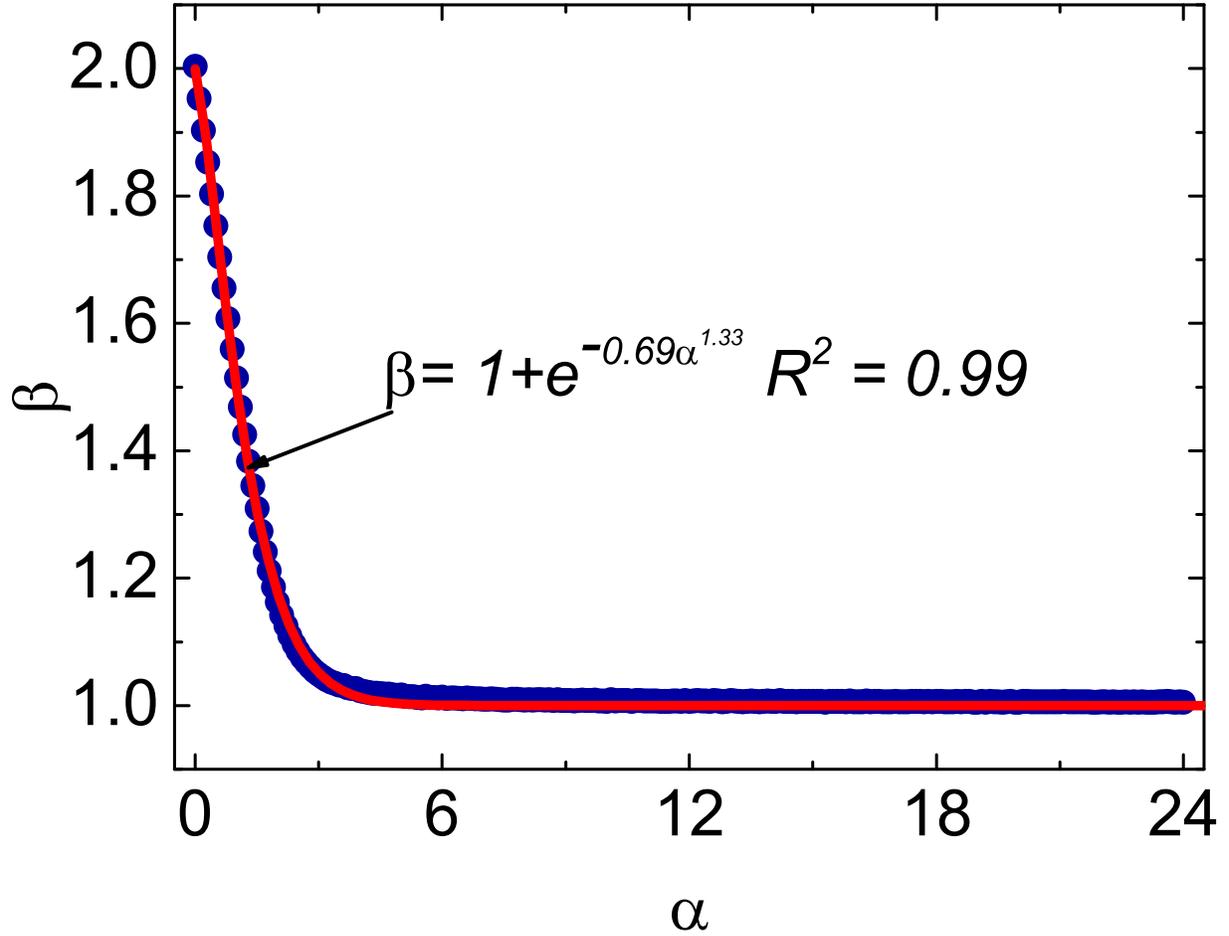}
% Here is how to import EPS art
\caption{The plot of $\beta$ as the function of $\alpha$. The horizontal space for adjacent two point is $0.1$. Each points denote the average value of 100 realizations of our model at $1000 \times 1000$ space (The total number of nodes in our network is $10000$). When $\alpha$ increases from $0$ to $\infty$, $\beta$ decreases more and more slowly and finally converges to $1$. Moreover, for each $\alpha$, we can get a unique $\beta$. Therefore, we conclude that our model provides a reliable tool to reproduce all superlinear relation under density value.}
\label{fig:avsb}
\end{figure}

Fig. \ref{fig:avsb} shows the scaling exponent $\beta$ as a function of $\alpha$. The red line is the fitting curve of an exponential function. $\beta$ monotonically decreases as $\alpha$ increases. Notice that when $\alpha = 0$, the connecting probability from a new node to any node already in the network will be $P_{j \rightarrow i} = \frac{r_{ij}^{-0}}{r_{min}^{-0}} = 1.0$. It means a new node will connect to all other nodes, and the network is a complete network at each time step. And then, it will lead to $E(t) \sim N^{2}(t) = (\rho(t) * L^2)^{2} = \rho^{2}(t) * L^{4}$. When $\alpha$ increases, $P_{j \rightarrow i}$ will decrease, and cause $\beta$ to decline. When $\alpha \to \infty$, eq. (\ref{eq:con_p}) can be rewriten as
\begin{equation}\label{eq:a infi}
P_{j \to i} = \lim\limits_{\alpha \to \infty}(\frac{ r_{min} }{ r_{ij} })^{\alpha} =
\begin{cases}
1 & r_{min} = r_{ij},\\
0 & r_{min} < r_{ij}.
\end{cases}
\end{equation}
It means that a new node can only connect to its nearest neighbor(s). To be specific, the relation between $E$ and $\rho$ is shown as the equation $E(t) = \rho(t)* L^2$, since only one edge is formed when a new is introduced into the $L \times L$ space. In this condition, the superlinear exponent $\beta$ equals to $1$.
Therefore, the exponent $\beta$ derives from simulation for our model can cover the whole range of realistic superlinear relation exponent $\beta$ ($1 \leq \beta \leq 2$). Moreover, since $\alpha$ and $\beta$ follow a one-to-one relation and $\beta$ is almost stable for different realizations at given $\alpha$, we can get a unique $\beta$ for each $\alpha$. Therefore, we can conclude that we provide a reliable tool to reproduce any superlinear relations under density value crossing different regions.

\section{Conclusion}\label{sec:Discussion}
The closing decades of the last century has seen the inception of the rapid socioeconomic development. Human's connection and communication in different regions continuously provide fuel for the engine of regional development. Now, the rapid socioeconomic development has become an inexorable challenge for all countries around the world. From the purely search for the rule of population growth, like the Clark model \cite{Clark1951}, to the rank-size relation \cite{zipf1949human,Gabaix1999}, to Vthe mechanism of regional sprawl \cite{Batty1994,Makse1995,Makse1998}, scientists have made great progress in dealing with this challenge. Recently, the superlinear relation discovered in real regional systems makes people penetrate more into the nature of these systems.

In this paper, we revisit the superlinear relations under density and total value. The superlinear scalings under density value are stronger than those in total value, and the scaling exponent $\beta$ under density value for different regions is bigger than $1$, which are consistent with Pan's results \cite{Pan2013}. According to the investigation on six different regional levels, superlinear scaling is empirically proved to be a general feature for different regions.

Despite the general feature, empirical studies also show that the suplinear scaling under density value lies in a range (1, 2], which is wider than the prior work \cite{Pan2013}. Therefore, an alternative model based on 2D lattice space is provided to cover the wider range. Numerical studies proved that, when linear size $L \ge 400$, there is no size dependence in our model, and the generated scaling exponent $\beta$ is stably tunable by the only parameter $\alpha$.

Scaling exponent $\beta$ monotonically decreases with the increase of parameter $\alpha$. It means the decrease of the intensity of spatial correlation weakens the power of regional development. Take Northwest and Northeast China as an example (see Fig. S2). Because the developing stage of the communication and transportation infrastructures in Northwest China is relatively lower than that in Northeast China, the spatial correlation in Northwest China is much weaker than that in Northeast China. Therefore, the superlinear exponent $\beta$ of the Northwest part of is much smaller.

Even though our model provides a general platform for extensive urban and regional studies, further improvement is still needed. First, an analytic result for the relation between $\alpha$ and $\beta$ is still open. Second, our model reveals that the different spatial correlation factor $\alpha$ leads to different scaling exponents $\beta$, while the origin of $\alpha$ is unclear. In fact, many factors, such as the transportation convenience and the communication accessibility etc., may influence the intensity of the spatial correlation. %If more data on them are available, it will be helpful for us to understand the orgin of $\alpha$.

\section*{Acknowledgements}
We thank the financial support from the Major State Basic Research Development Program of China (973 Program) No. 2012CB725400, the National Natural Science Foundation of China (61304147, 71101009, 71131001). We thank Zengru Di (BNU), Jinshan Wu(BNU) for very useful discussion.

%\nocite{*}

%\bibliography{mypaper}% Produces the bibliography via BibTeX.

\begin{thebibliography}{29}%
\makeatletter
\providecommand \@ifxundefined [1]{%
 \@ifx{#1\undefined}
}%
\providecommand \@ifnum [1]{%
 \ifnum #1\expandafter \@firstoftwo
 \else \expandafter \@secondoftwo
 \fi
}%
\providecommand \@ifx [1]{%
 \ifx #1\expandafter \@firstoftwo
 \else \expandafter \@secondoftwo
 \fi
}%
\providecommand \natexlab [1]{#1}%
\providecommand \enquote  [1]{``#1''}%
\providecommand \bibnamefont  [1]{#1}%
\providecommand \bibfnamefont [1]{#1}%
\providecommand \citenamefont [1]{#1}%
\providecommand \href@noop [0]{\@secondoftwo}%
\providecommand \href [0]{\begingroup \@sanitize@url \@href}%
\providecommand \@href[1]{\@@startlink{#1}\@@href}%
\providecommand \@@href[1]{\endgroup#1\@@endlink}%
\providecommand \@sanitize@url [0]{\catcode `\\12\catcode `\$12\catcode
  `\&12\catcode `\#12\catcode `\^12\catcode `\_12\catcode `\%12\relax}%
\providecommand \@@startlink[1]{}%
\providecommand \@@endlink[0]{}%
\providecommand \url  [0]{\begingroup\@sanitize@url \@url }%
\providecommand \@url [1]{\endgroup\@href {#1}{\urlprefix }}%
\providecommand \urlprefix  [0]{URL }%
\providecommand \Eprint [0]{\href }%
\providecommand \doibase [0]{http://dx.doi.org/}%
\providecommand \selectlanguage [0]{\@gobble}%
\providecommand \bibinfo  [0]{\@secondoftwo}%
\providecommand \bibfield  [0]{\@secondoftwo}%
\providecommand \translation [1]{[#1]}%
\providecommand \BibitemOpen [0]{}%
\providecommand \bibitemStop [0]{}%
\providecommand \bibitemNoStop [0]{.\EOS\space}%
\providecommand \EOS [0]{\spacefactor3000\relax}%
\providecommand \BibitemShut  [1]{\csname bibitem#1\endcsname}%
\let\auto@bib@innerbib\@empty
%</preamble>
\bibitem [{\citenamefont {Bettencourt}\ and\ \citenamefont
  {West}(2010)}]{Bettencourt2010}%
  \BibitemOpen
  \bibfield  {author} {\bibinfo {author} {\bibfnamefont {L.~M.~A.}\
  \bibnamefont {Bettencourt}}\ and\ \bibinfo {author} {\bibfnamefont
  {G.}~\bibnamefont {West}},\ }\href {\doibase 10.1038/467912a} {\bibfield
  {journal} {\bibinfo  {journal} {Nature}\ }\textbf {\bibinfo {volume} {467}},\
  \bibinfo {pages} {912} (\bibinfo {year} {2010})}\BibitemShut {NoStop}%
\bibitem [{\citenamefont {Bettencourt}\ \emph
  {et~al.}(2007{\natexlab{a}})\citenamefont {Bettencourt}, \citenamefont
  {Lobo},\ and\ \citenamefont {Strumsky}}]{Bettencourt2007}%
  \BibitemOpen
  \bibfield  {author} {\bibinfo {author} {\bibfnamefont {L.~M.~A.}\
  \bibnamefont {Bettencourt}}, \bibinfo {author} {\bibfnamefont
  {J.}~\bibnamefont {Lobo}}, \ and\ \bibinfo {author} {\bibfnamefont
  {D.}~\bibnamefont {Strumsky}},\ }\href@noop {} {\bibfield  {journal}
  {\bibinfo  {journal} {Research Policy}\ }\textbf {\bibinfo {volume} {36}},\
  \bibinfo {pages} {107} (\bibinfo {year} {2007}{\natexlab{a}})}\BibitemShut
  {NoStop}%
\bibitem [{\citenamefont {Shalizi}(2011)}]{Shalizi2011}%
  \BibitemOpen
  \bibfield  {author} {\bibinfo {author} {\bibfnamefont {C.~R.}\ \bibnamefont
  {Shalizi}},\ }\href@noop {} {} (\bibinfo {year} {2011}),\ \Eprint
  {http://arxiv.org/abs/arXiv:1102.4101} {arXiv:1102.4101} \BibitemShut
  {NoStop}%
\bibitem [{\citenamefont {Ortman}\ \emph {et~al.}(2014)\citenamefont {Ortman},
  \citenamefont {Cabaniss}, \citenamefont {Sturm},\ and\ \citenamefont
  {Bettencourt}}]{Ortman2014}%
  \BibitemOpen
  \bibfield  {author} {\bibinfo {author} {\bibfnamefont {S.~G.}\ \bibnamefont
  {Ortman}}, \bibinfo {author} {\bibfnamefont {A.~H.}\ \bibnamefont
  {Cabaniss}}, \bibinfo {author} {\bibfnamefont {J.~O.}\ \bibnamefont {Sturm}},
  \ and\ \bibinfo {author} {\bibfnamefont {L.~M.}\ \bibnamefont
  {Bettencourt}},\ }\href@noop {} {\bibfield  {journal} {\bibinfo  {journal}
  {PLoS ONE}\ }\textbf {\bibinfo {volume} {9}},\ \bibinfo {pages} {e87902}
  (\bibinfo {year} {2014})}\BibitemShut {NoStop}%
\bibitem [{\citenamefont {Arcaute}\ \emph {et~al.}(2014)\citenamefont
  {Arcaute}, \citenamefont {Hatna}, \citenamefont {Ferguson}, \citenamefont
  {Youn}, \citenamefont {Johansson},\ and\ \citenamefont
  {Batty}}]{Arcaute2014}%
  \BibitemOpen
  \bibfield  {author} {\bibinfo {author} {\bibfnamefont {E.}~\bibnamefont
  {Arcaute}}, \bibinfo {author} {\bibfnamefont {E.}~\bibnamefont {Hatna}},
  \bibinfo {author} {\bibfnamefont {P.}~\bibnamefont {Ferguson}}, \bibinfo
  {author} {\bibfnamefont {H.}~\bibnamefont {Youn}}, \bibinfo {author}
  {\bibfnamefont {A.}~\bibnamefont {Johansson}}, \ and\ \bibinfo {author}
  {\bibfnamefont {M.}~\bibnamefont {Batty}},\ }\href {\doibase
  10.1098/rsif.2014.0745} {\bibfield  {journal} {\bibinfo  {journal} {Journal
  of The Royal Society Interface}\ }\textbf {\bibinfo {volume} {12}},\ \bibinfo
  {pages} {20140745} (\bibinfo {year} {2014})}\BibitemShut {NoStop}%
\bibitem [{\citenamefont {Crane}\ and\ \citenamefont
  {Kinzig}(2005)}]{Crane2005}%
  \BibitemOpen
  \bibfield  {author} {\bibinfo {author} {\bibfnamefont {P.}~\bibnamefont
  {Crane}}\ and\ \bibinfo {author} {\bibfnamefont {A.}~\bibnamefont {Kinzig}},\
  }\href {\doibase 10.1126/science.1114165} {\bibfield  {journal} {\bibinfo
  {journal} {Science}\ }\textbf {\bibinfo {volume} {308}},\ \bibinfo {pages}
  {1225} (\bibinfo {year} {2005})}\BibitemShut {NoStop}%
\bibitem [{\citenamefont {Gonzalez}\ \emph {et~al.}(2008)\citenamefont
  {Gonzalez}, \citenamefont {Hidalgo},\ and\ \citenamefont
  {Barabasi}}]{Gonzalez2008}%
  \BibitemOpen
  \bibfield  {author} {\bibinfo {author} {\bibfnamefont {M.~C.}\ \bibnamefont
  {Gonzalez}}, \bibinfo {author} {\bibfnamefont {C.~A.}\ \bibnamefont
  {Hidalgo}}, \ and\ \bibinfo {author} {\bibfnamefont {A.-L.}\ \bibnamefont
  {Barabasi}},\ }\href@noop {} {\bibfield  {journal} {\bibinfo  {journal}
  {Nature}\ }\textbf {\bibinfo {volume} {453}},\ \bibinfo {pages} {779}
  (\bibinfo {year} {2008})}\BibitemShut {NoStop}%
\bibitem [{\citenamefont {Florida}(2002)}]{Florida2002}%
  \BibitemOpen
  \bibfield  {author} {\bibinfo {author} {\bibfnamefont {R.}~\bibnamefont
  {Florida}},\ }\href@noop {} {\bibfield  {journal} {\bibinfo  {journal} {The
  Washington Monthly}\ }\textbf {\bibinfo {volume} {34}},\ \bibinfo {pages}
  {15} (\bibinfo {year} {2002})}\BibitemShut {NoStop}%
\bibitem [{\citenamefont {Bettencourt}\ \emph
  {et~al.}(2007{\natexlab{b}})\citenamefont {Bettencourt}, \citenamefont
  {Lobo}, \citenamefont {Helbing}, \citenamefont {K¨¹hnert},\ and\
  \citenamefont {West}}]{Bettencourt2007a}%
  \BibitemOpen
  \bibfield  {author} {\bibinfo {author} {\bibfnamefont {L.~M.~A.}\
  \bibnamefont {Bettencourt}}, \bibinfo {author} {\bibfnamefont
  {J.}~\bibnamefont {Lobo}}, \bibinfo {author} {\bibfnamefont {D.}~\bibnamefont
  {Helbing}}, \bibinfo {author} {\bibfnamefont {C.}~\bibnamefont {K¨¹hnert}}, \
  and\ \bibinfo {author} {\bibfnamefont {G.~B.}\ \bibnamefont {West}},\ }\href
  {\doibase 10.1073/pnas.0610172104} {\bibfield  {journal} {\bibinfo  {journal}
  {Proc. Natl. Acad. Sci. U.S.A.}\ }\textbf {\bibinfo {volume} {104}},\
  \bibinfo {pages} {7301} (\bibinfo {year} {2007}{\natexlab{b}})}\BibitemShut
  {NoStop}%
\bibitem [{\citenamefont {Bettencourt}\ \emph {et~al.}(2010)\citenamefont
  {Bettencourt}, \citenamefont {Lobo}, \citenamefont {Strumsky},\ and\
  \citenamefont {West}}]{Bettencourt2010a}%
  \BibitemOpen
  \bibfield  {author} {\bibinfo {author} {\bibfnamefont {L.~M.~A.}\
  \bibnamefont {Bettencourt}}, \bibinfo {author} {\bibfnamefont
  {J.}~\bibnamefont {Lobo}}, \bibinfo {author} {\bibfnamefont {D.}~\bibnamefont
  {Strumsky}}, \ and\ \bibinfo {author} {\bibfnamefont {G.~B.}\ \bibnamefont
  {West}},\ }\href@noop {} {\bibfield  {journal} {\bibinfo  {journal} {PLoS
  ONE}\ }\textbf {\bibinfo {volume} {5}},\ \bibinfo {pages} {e13541} (\bibinfo
  {year} {2010})}\BibitemShut {NoStop}%
\bibitem [{\citenamefont {Bettencourt}(2013)}]{Bettencourt2013}%
  \BibitemOpen
  \bibfield  {author} {\bibinfo {author} {\bibfnamefont {L.~M.~A.}\
  \bibnamefont {Bettencourt}},\ }\href@noop {} {\bibfield  {journal} {\bibinfo
  {journal} {Science}\ }\textbf {\bibinfo {volume} {340}},\ \bibinfo {pages}
  {1438} (\bibinfo {year} {2013})}\BibitemShut {NoStop}%
\bibitem [{\citenamefont {Arbesman}\ \emph {et~al.}(2009)\citenamefont
  {Arbesman}, \citenamefont {Kleinberg},\ and\ \citenamefont
  {Strogatz}}]{Arbesman2009}%
  \BibitemOpen
  \bibfield  {author} {\bibinfo {author} {\bibfnamefont {S.}~\bibnamefont
  {Arbesman}}, \bibinfo {author} {\bibfnamefont {J.~M.}\ \bibnamefont
  {Kleinberg}}, \ and\ \bibinfo {author} {\bibfnamefont {S.~H.}\ \bibnamefont
  {Strogatz}},\ }\href {\doibase 10.1103/PhysRevE.79.016115} {\bibfield
  {journal} {\bibinfo  {journal} {Phys. Rev. E}\ }\textbf {\bibinfo {volume}
  {79}},\ \bibinfo {pages} {016115} (\bibinfo {year} {2009})}\BibitemShut
  {NoStop}%
\bibitem [{\citenamefont {Ahn}\ \emph {et~al.}(2010)\citenamefont {Ahn},
  \citenamefont {Bagrow},\ and\ \citenamefont {Lehmann}}]{Ahn2010}%
  \BibitemOpen
  \bibfield  {author} {\bibinfo {author} {\bibfnamefont {Y.-Y.}\ \bibnamefont
  {Ahn}}, \bibinfo {author} {\bibfnamefont {J.~P.}\ \bibnamefont {Bagrow}}, \
  and\ \bibinfo {author} {\bibfnamefont {S.}~\bibnamefont {Lehmann}},\
  }\href@noop {} {\bibfield  {journal} {\bibinfo  {journal} {Nature}\ }\textbf
  {\bibinfo {volume} {466}},\ \bibinfo {pages} {761} (\bibinfo {year}
  {2010})}\BibitemShut {NoStop}%
\bibitem [{\citenamefont {Mucha}\ \emph {et~al.}(2010)\citenamefont {Mucha},
  \citenamefont {Richardson}, \citenamefont {Macon}, \citenamefont {Porter},\
  and\ \citenamefont {Onnela}}]{Mucha2010}%
  \BibitemOpen
  \bibfield  {author} {\bibinfo {author} {\bibfnamefont {P.~J.}\ \bibnamefont
  {Mucha}}, \bibinfo {author} {\bibfnamefont {T.}~\bibnamefont {Richardson}},
  \bibinfo {author} {\bibfnamefont {K.}~\bibnamefont {Macon}}, \bibinfo
  {author} {\bibfnamefont {M.~A.}\ \bibnamefont {Porter}}, \ and\ \bibinfo
  {author} {\bibfnamefont {J.-P.}\ \bibnamefont {Onnela}},\ }\href@noop {}
  {\bibfield  {journal} {\bibinfo  {journal} {Science}\ }\textbf {\bibinfo
  {volume} {328}},\ \bibinfo {pages} {876} (\bibinfo {year}
  {2010})}\BibitemShut {NoStop}%
\bibitem [{\citenamefont {Onnela}\ \emph {et~al.}(2011)\citenamefont {Onnela},
  \citenamefont {Arbesman}, \citenamefont {Gonz{\'a}lez}, \citenamefont
  {Barab{\'a}si},\ and\ \citenamefont {Christakis}}]{Onnela2011}%
  \BibitemOpen
  \bibfield  {author} {\bibinfo {author} {\bibfnamefont {J.-P.}\ \bibnamefont
  {Onnela}}, \bibinfo {author} {\bibfnamefont {S.}~\bibnamefont {Arbesman}},
  \bibinfo {author} {\bibfnamefont {M.~C.}\ \bibnamefont {Gonz{\'a}lez}},
  \bibinfo {author} {\bibfnamefont {A.-L.}\ \bibnamefont {Barab{\'a}si}}, \
  and\ \bibinfo {author} {\bibfnamefont {N.~A.}\ \bibnamefont {Christakis}},\
  }\href@noop {} {\bibfield  {journal} {\bibinfo  {journal} {PLoS ONE}\
  }\textbf {\bibinfo {volume} {6}},\ \bibinfo {pages} {e16939} (\bibinfo {year}
  {2011})}\BibitemShut {NoStop}%
\bibitem [{\citenamefont {Expert}\ \emph {et~al.}(2011)\citenamefont {Expert},
  \citenamefont {Evans}, \citenamefont {Blondel},\ and\ \citenamefont
  {Lambiotte}}]{Expert2011}%
  \BibitemOpen
  \bibfield  {author} {\bibinfo {author} {\bibfnamefont {P.}~\bibnamefont
  {Expert}}, \bibinfo {author} {\bibfnamefont {T.~S.}\ \bibnamefont {Evans}},
  \bibinfo {author} {\bibfnamefont {V.~D.}\ \bibnamefont {Blondel}}, \ and\
  \bibinfo {author} {\bibfnamefont {R.}~\bibnamefont {Lambiotte}},\ }\href@noop
  {} {\bibfield  {journal} {\bibinfo  {journal} {Proc. Natl. Acad. Sci.
  U.S.A.}\ }\textbf {\bibinfo {volume} {108}},\ \bibinfo {pages} {7663}
  (\bibinfo {year} {2011})}\BibitemShut {NoStop}%
\bibitem [{\citenamefont {Yakubo}\ \emph {et~al.}(2014)\citenamefont {Yakubo},
  \citenamefont {Saijo},\ and\ \citenamefont {Koro{\v{s}}ak}}]{Yakubo2014}%
  \BibitemOpen
  \bibfield  {author} {\bibinfo {author} {\bibfnamefont {K.}~\bibnamefont
  {Yakubo}}, \bibinfo {author} {\bibfnamefont {Y.}~\bibnamefont {Saijo}}, \
  and\ \bibinfo {author} {\bibfnamefont {D.}~\bibnamefont {Koro{\v{s}}ak}},\
  }\href@noop {} {\bibfield  {journal} {\bibinfo  {journal} {Phys. Rev. E}\
  }\textbf {\bibinfo {volume} {90}},\ \bibinfo {pages} {022803} (\bibinfo
  {year} {2014})}\BibitemShut {NoStop}%
\bibitem [{\citenamefont {Pan}\ \emph {et~al.}(2013)\citenamefont {Pan},
  \citenamefont {Ghoshal}, \citenamefont {Krumme}, \citenamefont {Cebrian},\
  and\ \citenamefont {Pentland}}]{Pan2013}%
  \BibitemOpen
  \bibfield  {author} {\bibinfo {author} {\bibfnamefont {W.}~\bibnamefont
  {Pan}}, \bibinfo {author} {\bibfnamefont {G.}~\bibnamefont {Ghoshal}},
  \bibinfo {author} {\bibfnamefont {C.}~\bibnamefont {Krumme}}, \bibinfo
  {author} {\bibfnamefont {M.}~\bibnamefont {Cebrian}}, \ and\ \bibinfo
  {author} {\bibfnamefont {A.}~\bibnamefont {Pentland}},\ }\href {\doibase
  10.1038/ncomms2961} {\bibfield  {journal} {\bibinfo  {journal} {Nat Commun}\
  }\textbf {\bibinfo {volume} {4}},\ \bibinfo {pages} {1961} (\bibinfo {year}
  {2013})}\BibitemShut {NoStop}%
\bibitem [{\citenamefont {Pumain}\ \emph {et~al.}(2006)\citenamefont {Pumain},
  \citenamefont {Paulus}, \citenamefont {Vacchiani-Marcuzzo},\ and\
  \citenamefont {Lobo}}]{Pumain2006}%
  \BibitemOpen
  \bibfield  {author} {\bibinfo {author} {\bibfnamefont {D.}~\bibnamefont
  {Pumain}}, \bibinfo {author} {\bibfnamefont {F.}~\bibnamefont {Paulus}},
  \bibinfo {author} {\bibfnamefont {C.}~\bibnamefont {Vacchiani-Marcuzzo}}, \
  and\ \bibinfo {author} {\bibfnamefont {J.}~\bibnamefont {Lobo}},\ }\href@noop
  {} {\bibfield  {journal} {\bibinfo  {journal} {Cybergeo : European Journal of
  Geography}\ }\textbf {\bibinfo {volume} {2006}},\ \bibinfo {pages} {343}
  (\bibinfo {year} {2006})}\BibitemShut {NoStop}%
\bibitem [{\citenamefont {Nomaler}\ \emph {et~al.}(2014)\citenamefont
  {Nomaler}, \citenamefont {Frenken},\ and\ \citenamefont
  {Heimeriks}}]{Nomaler2014}%
  \BibitemOpen
  \bibfield  {author} {\bibinfo {author} {\bibfnamefont {{\"O}.}~\bibnamefont
  {Nomaler}}, \bibinfo {author} {\bibfnamefont {K.}~\bibnamefont {Frenken}}, \
  and\ \bibinfo {author} {\bibfnamefont {G.}~\bibnamefont {Heimeriks}},\
  }\href@noop {} {\bibfield  {journal} {\bibinfo  {journal} {PLoS ONE}\
  }\textbf {\bibinfo {volume} {9}},\ \bibinfo {pages} {e110805} (\bibinfo
  {year} {2014})}\BibitemShut {NoStop}%
\bibitem [{\citenamefont {Zhang}\ and\ \citenamefont {Yu}(2010)}]{Zhang2010}%
  \BibitemOpen
  \bibfield  {author} {\bibinfo {author} {\bibfnamefont {J.}~\bibnamefont
  {Zhang}}\ and\ \bibinfo {author} {\bibfnamefont {T.}~\bibnamefont {Yu}},\
  }\href@noop {} {\bibfield  {journal} {\bibinfo  {journal} {Physica A}\
  }\textbf {\bibinfo {volume} {389}},\ \bibinfo {pages} {4887} (\bibinfo {year}
  {2010})}\BibitemShut {NoStop}%
\bibitem [{\citenamefont {Alves}\ \emph {et~al.}(2014)\citenamefont {Alves},
  \citenamefont {Ribeiro}, \citenamefont {Lenzi},\ and\ \citenamefont
  {Mendes}}]{Alves2014}%
  \BibitemOpen
  \bibfield  {author} {\bibinfo {author} {\bibfnamefont {L.}~\bibnamefont
  {Alves}}, \bibinfo {author} {\bibfnamefont {H.}~\bibnamefont {Ribeiro}},
  \bibinfo {author} {\bibfnamefont {E.}~\bibnamefont {Lenzi}}, \ and\ \bibinfo
  {author} {\bibfnamefont {R.}~\bibnamefont {Mendes}},\ }\href@noop {}
  {\bibfield  {journal} {\bibinfo  {journal} {Physica A}\ }\textbf {\bibinfo
  {volume} {409}},\ \bibinfo {pages} {175} (\bibinfo {year}
  {2014})}\BibitemShut {NoStop}%
\bibitem [{\citenamefont {Kleinberg}(2000)}]{Kleinberg2000}%
  \BibitemOpen
  \bibfield  {author} {\bibinfo {author} {\bibfnamefont {J.~M.}\ \bibnamefont
  {Kleinberg}},\ }\href {\doibase 10.1038/35022643} {\bibfield  {journal}
  {\bibinfo  {journal} {Nature}\ }\textbf {\bibinfo {volume} {406}},\ \bibinfo
  {pages} {845} (\bibinfo {year} {2000})}\BibitemShut {NoStop}%
\bibitem [{\citenamefont {Clark}(1951)}]{Clark1951}%
  \BibitemOpen
  \bibfield  {author} {\bibinfo {author} {\bibfnamefont {C.}~\bibnamefont
  {Clark}},\ }\href@noop {} {\bibfield  {journal} {\bibinfo  {journal} {J R
  Stat Soc A Stat.}\ }\textbf {\bibinfo {volume} {114}},\ \bibinfo {pages}
  {490} (\bibinfo {year} {1951})}\BibitemShut {NoStop}%
\bibitem [{\citenamefont {Zipf}(1949)}]{zipf1949human}%
  \BibitemOpen
  \bibfield  {author} {\bibinfo {author} {\bibfnamefont {G.~K.}\ \bibnamefont
  {Zipf}},\ }\href@noop {} {\emph {\bibinfo {title} {Human behavior and the
  principle of least effort.}}}\ (\bibinfo  {publisher} {addison-wesley
  press},\ \bibinfo {year} {1949})\BibitemShut {NoStop}%
\bibitem [{\citenamefont {Gabaix}(1999)}]{Gabaix1999}%
  \BibitemOpen
  \bibfield  {author} {\bibinfo {author} {\bibfnamefont {X.}~\bibnamefont
  {Gabaix}},\ }\href@noop {} {\bibfield  {journal} {\bibinfo  {journal} {The
  Quarterly journal of economics}\ }\textbf {\bibinfo {volume} {114}},\
  \bibinfo {pages} {739} (\bibinfo {year} {1999})}\BibitemShut {NoStop}%
\bibitem [{\citenamefont {Batty}\ and\ \citenamefont
  {Longley}(1994)}]{Batty1994}%
  \BibitemOpen
  \bibfield  {author} {\bibinfo {author} {\bibfnamefont {M.}~\bibnamefont
  {Batty}}\ and\ \bibinfo {author} {\bibfnamefont {P.}~\bibnamefont
  {Longley}},\ }\href@noop {} {\emph {\bibinfo {title} {Fractal Cities}}}\
  (\bibinfo  {publisher} {Academic, San Diego},\ \bibinfo {year}
  {1994})\BibitemShut {NoStop}%
\bibitem [{\citenamefont {Makse}\ \emph {et~al.}(1995)\citenamefont {Makse},
  \citenamefont {Havlin},\ and\ \citenamefont {Stanley}}]{Makse1995}%
  \BibitemOpen
  \bibfield  {author} {\bibinfo {author} {\bibfnamefont {H.~A.}\ \bibnamefont
  {Makse}}, \bibinfo {author} {\bibfnamefont {S.}~\bibnamefont {Havlin}}, \
  and\ \bibinfo {author} {\bibfnamefont {H.~E.}\ \bibnamefont {Stanley}},\
  }\href@noop {} {\bibfield  {journal} {\bibinfo  {journal} {Nature}\ }\textbf
  {\bibinfo {volume} {377}},\ \bibinfo {pages} {608 } (\bibinfo {year}
  {1995})}\BibitemShut {NoStop}%
\bibitem [{\citenamefont {Makse}\ \emph {et~al.}(1998)\citenamefont {Makse},
  \citenamefont {Andrade}, \citenamefont {Batty}, \citenamefont {Havlin},\ and\
  \citenamefont {Stanley}}]{Makse1998}%
  \BibitemOpen
  \bibfield  {author} {\bibinfo {author} {\bibfnamefont {H.~A.}\ \bibnamefont
  {Makse}}, \bibinfo {author} {\bibfnamefont {J.~S.}\ \bibnamefont {Andrade}},
  \bibinfo {author} {\bibfnamefont {M.}~\bibnamefont {Batty}}, \bibinfo
  {author} {\bibfnamefont {S.}~\bibnamefont {Havlin}}, \ and\ \bibinfo {author}
  {\bibfnamefont {H.~E.}\ \bibnamefont {Stanley}},\ }\href {\doibase
  10.1103/PhysRevE.58.7054} {\bibfield  {journal} {\bibinfo  {journal} {Phys.
  Rev. E}\ }\textbf {\bibinfo {volume} {58}},\ \bibinfo {pages} {7054}
  (\bibinfo {year} {1998})}\BibitemShut {NoStop}%
\end{thebibliography}

%

\newpage

\begin{table}
\caption{The comparison of scaling relations between density value and total value for different regions in China (see Fig. S1). The adjusted $R$ square (adj-$R^2$) shows that the relation between $GDP_{density}$ and $POP_{density}$ fits much better than the relation between GDP and POP. Moreover, the scaling exponents $\beta$ under density value are bigger than $1$, which is consistent with Pan's result. The scaling exponents under density value, which are out of the range [1.1, 1.3], are shown in bold.}\label{tab:Chinese_Regions}
\begin{ruledtabular}
\begin{tabular}{ccccc}
  \multirow{2}{*}{Regions in China} &\multicolumn{2}{l}{GDP vs. POP}&\multicolumn{2}{l}{GDP$_{density}$ vs. POP$_{density}$}\\ \cline{2-3} \cline{4-5}
  & $\beta$ & adj-$R^2$ & $\beta$ & adj-$R^2$\\ \hline
  Country & 1.03 & 0.67 & 1.11 & 0.89 \\
  East & 0.93 & {0.53} & 1.21 & 0.72\\
  North & 0.91 & {0.56} & \textbf{1.02} & 0.86\\
  Northeast & 0.82 & {0.54} & \textbf{1.33} & 0.87\\
  Northwest & 0.85 & {0.50} & \textbf{1.00(3)} & 0.83\\
  South Central & 0.87 & {0.47}& 1.29 & 0.86\\
  Southwest & 1.05 & {0.87} & \textbf{1.04} & 0.95\\
\end{tabular}
\end{ruledtabular}
\end{table}

\begin{table}
\caption{Classification of 6 levels of geographic region. All of scaling exponents are bigger than $1$, which means superlinear relation under density value widely exists at different regions. The scaling exponents, which are out of the range [1.1, 1.3], are also shown in bold.}\label{tab:region_levels}
%\begin{ruledtabular}
\begin{tabular}{lclc}
  \hline
  \hline
  Level & Count Element* & Example & $\beta$\\
  \hline
  World & Country & -- & 1.14 $\pm$ 0.15\\
  \hline
  Continent & Country & Asia & 1.13 $\pm$ 0.15\\
   &  & Europe & $1.22 \pm0.15$\\
  \hline
  Country & City & China & 1.11 $\pm$ 0.02 \\
   & & USA & \textbf{1.08} $\pm$ 0.01\\
   & & Japan & \textbf{1.08} $\pm$ 0.02\\
  \hline
  Region & City & Northeast China & \textbf{1.33} $\pm$ 0.09\\
  \hline
  Province& County & Guangdong & 1.24 $\pm$ 0.05\\
   & & Heilongjiang & \textbf{1.03} $\pm$ 0.06\\
   & & Henan & 1.16 $\pm$ 0.06\\
  \hline
  City& District & Hangzhou & 1.15 $\pm$ 0.06 \\
   & & Chongqing & \textbf{1.42} $\pm$ 0.04 \\
   & & Xian & \textbf{1.31} $\pm$ 0.05\\
  \hline
  \hline
\end{tabular}
\begin{tablenotes}
\centering
\item \scriptsize*City is a bigger administrative unit than County in China
\end{tablenotes}
%\end{ruledtabular}
\end{table}

\end{document}